\documentclass[%
  reprint,
  letterpaper,
  superscriptaddress,
  groupedaddress,
  showpacs,preprintnumbers,
  amsmath,amssymb,
  aps,
  prl,
]{revtex4-1}

\usepackage[pdftex]{graphicx}
\usepackage{dcolumn}
\usepackage{bm}

\begin{document}


\title{Growth fluctuation in preferential attachment dynamics}

\author{Yasuhiro Hashimoto}
\email[]{hashi@cs.tsukuba.ac.jp}
\affiliation{Division of Information Engineering,\\
  Faculty of Engineering, Information and Systems,\\
  University of Tsukuba%
}

\date{\today}

\begin{abstract}
  In the Yule--Simon process, selection of words follows the {\it preferential attachment} mechanism,
  resulting in the power-law growth in the cumulative number of individual word occurrences.
  This is derived using mean-field approximation,
  assuming a continuum limit of both the time and number of word occurrences.
  However, time and word occurrences are inherently discrete in the process,
  and it is natural to assume that the cumulative number of word occurrences has a certain fluctuation
  around the average behavior predicted by the mean-field approximation.
  We derive the exact and approximate forms of the probability distribution of such fluctuation analytically
  and confirm that those probability distributions are well supported by the numerical experiments.
\end{abstract}

\pacs{02.50.Ey, 05.40.-a, 87.23.Ge, 89.75.Hc}

\maketitle

The Yule--Simon process is a classical mathematical model
that describes a branching process in discrete time and state space;
it was originally introduced by Yule to explain the population dynamics of biological species
in continuous time and discrete state space \cite{Yule1924,Bacaer2011,Simkin2011}
and later modified by Simon into the discrete time and state model \cite{Simon1955,Simkin2011}.
In Simon's scheme, the process yields a {\it word} sequence.
A word is added to the sequence at every time step,
where a new word, or vocabulary, is created with probability \(\alpha\),
whereas with the complementary probability \(1-\alpha\), or \(\bar{\alpha}\),
one of the existing words in the sequence is chosen again.
This process is analogous to that of book reading,
where novel or known words appear one after another sequentially.
One of the significant results of Yule's and Simon's works is the derivation of the population distribution
that follows the power-law form, also known as Zipf's law in the rank-frequency distribution \cite{Zipf1935}.

Now following Simon's scheme,
let us denote \(i\) as the index of distinct words sorted in the ascending order of time when they are created.
The probability of word \(i\) being chosen among the existing words is proportional to
the number of occurrences of word \(i\) in the sequence, and this is defined as follows:
\begin{align}
  P(i,t)=n_i(t)/N(t),
  \label{eq:preferential_attachment}
\end{align}
where \(n_i(t)\) is the cumulative number of occurrences of word \(i\) until time step \(t\)
and \(N(t)\) is the length of the sequence at \(t\), that is,
the total number of word occurrences until \(t\)---\(N(t)=t\) from the definition.
The name of the {\it preferential attachment} mechanism
derives from this proportionality in the word selection,
sharing the same idea as the well-known {\it urn} models \cite{Mahmoud2008}.
We should note that Simon himself assumed rather a weaker condition
than that in Eq.~(\ref{eq:preferential_attachment}),
which is equivalent to that implicitly assumed in Yule's scheme.
Instead, Simon introduced the notion of {\it class},
a group of distinct words of the same number of occurrences,
to be chosen in proportion to the size of the class,
that is, the total number of word occurrences included in the class;
meanwhile, the rule determining which word is actually picked up in the chosen class is arbitrary.
Thus, the probability of the class being chosen is defined as follows:
\begin{align}
  \mathcal{P}(n,t)=nf(n,t)/N(t),
  \label{eq:preferential_attachment_simon}
\end{align}
where \(n\) is the cumulative number of word occurrences, or the class, and
\(f(n,t)\) is the number of distinct words included in class \(n\) at time \(t\).
If we adopt the additional rule to Eq.~(\ref{eq:preferential_attachment_simon})
that picks up a word uniformly at random in the chosen class,
it leads to the same result as in Eq.~(\ref{eq:preferential_attachment}).
We use the term ``the Yule--Simon process'' to refer to Eq.~(\ref{eq:preferential_attachment}),
and our study is based on this.

The Yule--Simon process has been used as an archetype of various other dynamic processes
such as the Barab\'{a}si--Albert (BA) graph model \cite{Barabasi1999},
which describes the growth of the web,
representing a specific case of the process when \(\alpha=1/2\).
In the BA graph, the graph grows by adding nodes (webpages) to the graph one by one,
resulting in a certain number of edges (hyperlinks) connected to the existing nodes
in proportion to their degree, that is, the number of edges belonging to the target node.
We see a direct correspondence between the models;
``node'' and ``degree'' appearing in the BA graph are paraphrases of
``word'' and ``word occurrence,'' respectively, in the Yule--Simon process \cite{Bornholdt2001}.
Barab\'{a}si and others analyzed how the node gathers the number of edges in the evolution
and showed that the degree grows in a power-law fashion
in the continuum limit of time and degree as follows:
\begin{align}
  k_i(t)\propto (t/t_i)^{1/2},
  \label{eq:expected_value_BA}
\end{align}
where \(k_i(t)\) is the expected degree of node \(i\) at time \(t\)
and \(t_i\) is the time when node \(i\) joined the graph.
Following the same logic,
the expected value of the cumulative number of occurrences of word \(i\) at time \(t\),
denoted by \(n^*_i(t)\), is derived as follows:
\begin{align*}
  n^*_i(t+\Delta t)=n^*_i(t)+(1-\alpha)P(i,t)\Delta t.
\end{align*}
Then, via the integral form
\begin{align}
  \int\frac{dn^*_i}{n^*_i}=(1-\alpha)\int\frac{dt}{t},
\end{align}
we obtain
\begin{align}
  n^*_i(t)=(t/t_i)^{1-\alpha}
  \label{eq:expected_value_YS}
\end{align}
using the initial condition \(n^*_i(t_i)=1\).
The homology between Eq.~(\ref{eq:expected_value_BA}) and Eq.~(\ref{eq:expected_value_YS}) implies
that the BA graph is actually a particular case of the Yule--Simon process with \(\alpha=1/2\).

The mean-field approximation elucidates the expected behavior
of the increase in the cumulative number of word occurrences
under the preferential attachment mechanism, as shown above.
Even so, we can assume that the individual word occurrence
will deviate from the expected value under a certain period of observation;
there might be words that occur more frequently than expected and others that appear less frequently.
We can likely attribute such individuality to
factors such as the so-called {\it fitness} \cite{Bianconi2001} of each word,
environmental contingency, or the inherent dynamics of the system.
What shape the probability distribution of such fluctuation has is an interesting question,
since anomalous behavior often attracts our interest more than ordinary behavior \cite{Krapivsky2002b};
in addition, knowing the shape of the distribution function might provide a useful theoretical baseline
to compare the growth of distinct words in, for example, social annotation systems \cite{Cattuto2009,Gupta2010}
and network elements in complex networks \cite{Albert2002}
that joined the system at close points in time.

Based on a similar motivation,
Krapivsky and Redner investigated the fluctuation of the degree distribution in networks,
that is, the fluctuation of the numbers of nodes that have the same degree \cite{Krapivsky2002a}.
In other examples, specific scaling laws between the {\it growth rate},
that is, the ratio of the sizes of system components at two consecutive time points,
and its fluctuation have been investigated in various social systems
such as city size, scientific output, human communication, and so on \cite{Gabaix1999,Matia2005,Rybski2009}.
Those works focus on the growth fluctuation of the class mentioned above,
that is, a group of system components that have the same size, as a function of each size.
In contrast, we focus on the fluctuation observed in individuality.

In the following, first we derive the probability distribution of the growth fluctuation
that the individual words exhibit under the preferential attachment mechanism analytically.
Then, we check the validity of the formula through a comparison with the results from numerical experiments.

Let us denote \(P\bigl(n_i(t)=n\bigr)\) as the probability of
the cumulative number of occurrences of word \(i\) at \(t\), denoted by \(n_i(t)\), to be equal to \(n\),
and \(P\bigl(n_i(t)\rightarrow n\bigr)\) as the probability of \(n_i(t)\)
to become \(n\) from \(n-1\) right at \(t\).
Introducing \(\tau\), an elapsed time from \(t_i\),
and \(s_i=t_i+\tau\) as the time to measure the probabilities,
\(P\bigl(n_i(s_i)=n\bigr)\) and \(P\bigl(n_i(s_i)\rightarrow n\bigr)\)
can be written recursively as follows: For \(n=1\),
\begin{align}
  P\bigl(n_i(s_i)=1\bigr)
  &=\prod_{t=t_i}^{s_i-1}\left(\alpha+\bar{\alpha}\frac{t-1}{t}\right)\nonumber\\
  &=\frac{\Gamma(t_i)\Gamma(s_i-\bar{\alpha})}{\Gamma(s_i)\Gamma(t_i-\bar{\alpha})},
  \label{eq:pn_1}
\end{align}
\begin{widetext}
and for \(n=2\),
\begin{align}
  P\bigl(n_i(s_i)\rightarrow2\bigr)
  &=P\bigl(n_i(s_i-1)=1\bigr)\frac{\bar{\alpha}}{s_i-1}\nonumber\\
  &=\bar{\alpha}\frac{\Gamma(t_i)\Gamma(s_i-1-\bar{\alpha})}{\Gamma(s_i)\Gamma(t_i-\bar{\alpha})},\nonumber\\
  P\bigl(n_i(s_i)=2\bigr)
  &=\sum_{u=t_i+1}^{s_i}\left[P\bigl(n_i(u)\rightarrow2\bigr)
    \prod_{t=u}^{s_i-1}\left(\alpha+\bar{\alpha}\frac{t-2}{t}\right)\right]\nonumber\\
  &=\bar{\alpha}\frac{\Gamma(t_i)\Gamma(s_i-2\bar{\alpha})}{\Gamma(s_i)\Gamma(t_i-\bar{\alpha})}
  \sum_{u=t_i+1}^{s_i}\frac{\Gamma(u-1-\bar{\alpha})}{\Gamma(u-2\bar{\alpha})}.
  \label{eq:pn_2}
\end{align}
\end{widetext}
Equation (\ref{eq:pn_1}) means that word \(i\) is not chosen for \(\tau\) since its first appearance.
Equation (\ref{eq:pn_2}) means that at a certain time point in the interval \([t_i+1:s_i]\),
word \(i\) is chosen only once and after that, it can never be chosen until \(s_i\).
Further, for \(n>2\), the form of the probabilities becomes more complicated
because it has the term of weighted and nested sums of the ratios of Gamma functions in it.
However, let us write down a few more values one by one:
\begin{widetext}
  \noindent For \(n=3\),
  \begin{align}
    P\bigl(n_i(s_i)\rightarrow3\bigr)
    &=P\bigl(n_i(s_i-1)=2\bigr)\frac{2\bar{\alpha}}{s_i-1}\nonumber\\
    &=2\bar{\alpha}^2\frac{\Gamma(t_i)\Gamma(s_i-1-2\bar{\alpha})}{\Gamma(s_i)\Gamma(t_i-\bar{\alpha})}
    \sum_{u=t_i+1}^{s_i-1}\frac{\Gamma(u-1-\bar{\alpha})}{\Gamma(u-2\bar{\alpha})},\nonumber\\
    P\bigl(n_i(s_i)=3\bigr)
    &=\sum_{u=t_i+2}^{s_i}\left[P\bigl(n_i(u)\rightarrow3\bigr)
      \prod_{t=u}^{s_i-1}\left(\alpha+\bar{\alpha}\frac{t-3}{t}\right)\right]\nonumber\\
    &=2\bar{\alpha}^2\frac{\Gamma(t_i)\Gamma(s_i-3\bar{\alpha})}{\Gamma(s_i)\Gamma(t_i-\bar{\alpha})}
    \sum_{u=t_i+2}^{s_i}\left[\frac{\Gamma(u-1-2\bar{\alpha})}{\Gamma(u-3\bar{\alpha})}
      \sum_{v=t_i+1}^{u-1}\frac{\Gamma(v-1-\bar{\alpha})}{\Gamma(v-2\bar{\alpha})}\right],
    \label{eq:pn_3}
  \end{align}
  and for \(n=4\),
  \begin{align}
    P\bigl(n_i(s_i)\rightarrow4\bigr)
    &=P\bigl(n_i(s_i-1)=3\bigr)\frac{3\bar{\alpha}}{s_i-1}\nonumber\\
    &=6\bar{\alpha}^3\frac{\Gamma(t_i)\Gamma(s_i-1-3\bar{\alpha})}{\Gamma(s_i)\Gamma(t_i-\bar{\alpha})}
    \sum_{u=t_i+2}^{s_i-1}\left[
      \frac{\Gamma(u-1-2\bar{\alpha})}{\Gamma(u-3\bar{\alpha})}
      \sum_{v=t_i+1}^{u-1}\frac{\Gamma(v-1-\bar{\alpha})}{\Gamma(v-2\bar{\alpha})}\right],\nonumber\\
    P\bigl(n_i(s_i)=4\bigr)
    &=\sum_{u=t_i+3}^{s_i}\left[P\bigl(n_i(u)\rightarrow4\bigr)
      \prod_{t=u}^{s_i-1}\left(\alpha+\bar{\alpha}\frac{t-4}{t}\right)\right]\nonumber\\
    &=6\bar{\alpha}^3\frac{\Gamma(t_i)\Gamma(s_i-4\bar{\alpha})}{\Gamma(s_i)\Gamma(t_i-\bar{\alpha})}
    \sum_{u=t_i+3}^{s_i}\left[\frac{\Gamma(u-1-3\bar{\alpha})}{\Gamma(u-4\bar{\alpha})}
      \sum_{v=t_i+2}^{u-1}\left[\frac{\Gamma(v-1-2\bar{\alpha})}{\Gamma(v-3\bar{\alpha})}
        \sum_{w=t_i+1}^{v-1}\frac{\Gamma(w-1-\bar{\alpha})}{\Gamma(w-2\bar{\alpha})}\right]\right].
    \label{eq:pn_4}
  \end{align}
\end{widetext}
Looking at Eqs.~(\ref{eq:pn_1}), (\ref{eq:pn_2}), (\ref{eq:pn_3}), and (\ref{eq:pn_4}) deliberately,
we can inductively infer their general form as follows:
\begin{align}
  P&\bigl(n_i(s_i)=n\bigr)=\nonumber\\
  &\begin{cases}
     \frac{\Gamma(t_i)\Gamma(s_i-\bar{\alpha})}{\Gamma(s_i)\Gamma(t_i-\bar{\alpha})} & \mbox{if \(n=1\)},\\
     (n-1)!\bar{\alpha}^{n-1}\frac{\Gamma(t_i)\Gamma(s_i-n\bar{\alpha})}{\Gamma(s_i)\Gamma(t_i-\bar{\alpha})}\sum_{\phi=t_i+n-1}^{s_i}\mathcal{S}_{n}(\phi) & \mbox{if \(n>1\)}.
   \end{cases}
  \label{eq:pn_formal}
\end{align}
The term \(\mathcal{S}_n(\phi)\) is defined as the following recursive function with a depth of \(n-1\):
\begin{align}
  \mathcal{S}_n&(\phi)=\nonumber\\
  &\begin{cases}
     \frac{\Gamma(\phi-1-\bar{\alpha})}{\Gamma(\phi-2\bar{\alpha})} & \mbox{if \(n=2\)},\\
     \frac{\Gamma(\phi-1-(n-1)\bar{\alpha})}{\Gamma(\phi-n\bar{\alpha})}\sum_{\psi=t_i+n-2}^{\phi-1}\mathcal{S}_{n-1}(\psi) & \mbox{if \(n> 2\)}.
   \end{cases}
  \label{eq:recursive_sum}
\end{align}
This is the exact form of the probability distribution wherein
the cumulative number of occurrences of word \(i\) at time \(s_i\) will be \(n\).
For sufficiently large values of \(t_i\) and \(s_i\),
these equations can be asymptotically transformed as follows:
\begin{align}
  P&\bigl(n_i(s_i)=n\bigr)\sim\nonumber\\
  &\begin{cases}
     t_i^{\bar{\alpha}}s_i^{-\bar{\alpha}} & \mbox{if \(n=1\)},\\
     (n-1)!\bar{\alpha}^{n-1}t_i^{\bar{\alpha}}s_i^{-n\bar{\alpha}}\sum_{\phi=t_i+n-1}^{s_i}\mathcal{S}_n(\phi) & \mbox{if \(n> 1\)},
   \end{cases}
  \label{eq:pn_formal_approx}
\end{align}
and
\begin{align}
  \mathcal{S}_n(\phi)
  \sim\begin{cases}
  \phi^{-\alpha} & \mbox{if \(n=2\)},\\
  \phi^{-\alpha}\sum_{\psi=t_i+n-2}^{\phi-1}\mathcal{S}_{n-1}(\psi) & \mbox{if \(n> 2\)},
  \end{cases}
  \label{eq:recursive_sum_approx}
\end{align}
where we use the asymptotic approximation of the ratio of Gamma functions for large \(t_i\);
\(\lim_{t\rightarrow\infty}\Gamma(t-a)/\Gamma(t)\sim t^{-a}\).
Equations (\ref{eq:pn_formal_approx}) and (\ref{eq:recursive_sum_approx}) represent
one of the principal results of this article.

If \(\alpha\rightarrow 0\), or \(\bar{\alpha}\rightarrow 1\),
all weighting factors \(\phi^{-\alpha}\) in Eq.~(\ref{eq:recursive_sum_approx}),
or all ratios of Gamma functions in Eq.~(\ref{eq:recursive_sum}), become exactly equal to 1.
Consequently, we obtain a specific value of the sum part of
Eqs.~(\ref{eq:pn_formal}) and (\ref{eq:pn_formal_approx}) as follows:
\begin{align}
  \sum_{\phi=t_i+n-1}^{s_i}\mathcal{S}_{n-1}(\phi)\underset{\alpha\rightarrow 0}{=}\frac{(\tau-n+2)^{n-1}}{(n-1)!},
  \label{eq:value_of_nested_sum}
\end{align}
which is the volume of an \((n-1)\)-dimensional triangular pyramid
where all of the edges aligned to a corresponding basis vector have the length \(\tau-n+2\).
Substituting Eq.~(\ref{eq:value_of_nested_sum}) into Eq.~(\ref{eq:pn_formal_approx}),
we obtain a relatively simple form, as follows:
\begin{align}
  P\bigl(n_i(s_i)=n\bigr)
  \underset{\alpha\rightarrow 0}{\sim}t_is_i^{-n}(\tau-n+2)^{n-1}.
  \label{eq:pn_alpha0}
\end{align}
 
Alternatively, in the case of larger \(\alpha\) such as \(1/2\) in the BA graph,
it is unclear whether a simple form like Eq.~(\ref{eq:pn_alpha0}) is available,
so that we have to numerically calculate
Eqs.~(\ref{eq:pn_formal_approx}) and (\ref{eq:recursive_sum_approx}) directly, if needed.
Practically, if we calculate all terms in the nested sum naively,
it requires approximately \(\tau^n\) operations, and such a large calculation will fail easily.
Once we calculate any of \(\mathcal{S}_m(\phi)\) (\(m\) starts from two),
storing and reusing the values associating with the pair of \(m\) and \(\phi\)
reduces the total amount of the calculation drastically, and will make the calculation feasible.

\begin{figure}
  \includegraphics[width=0.4\textwidth]{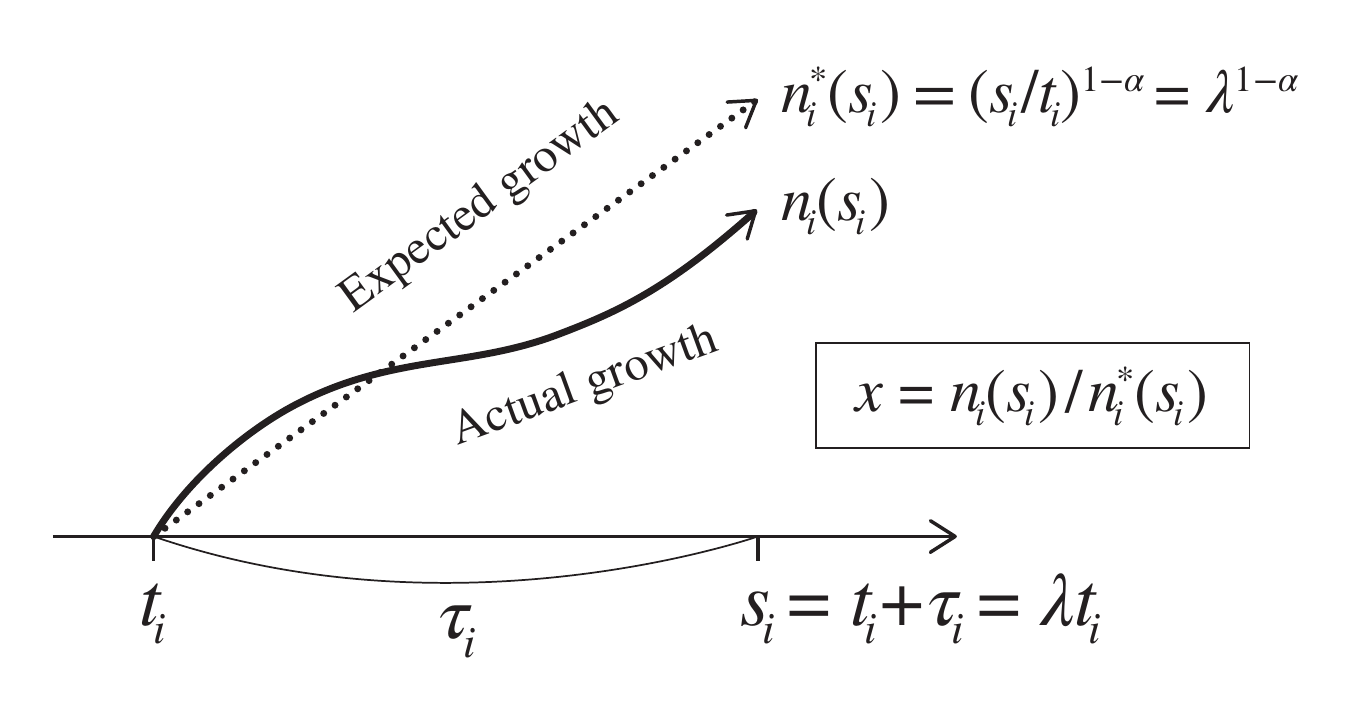}
  \caption{%
    A diagram of the relationships between the variables,
    depicting the growth of the cumulative number of word occurrences.
    }
  \label{fig:diagram}
\end{figure}

The following discussion is based on Eq.~(\ref{eq:pn_alpha0}),
the particular form for a sufficiently small \(\alpha\).
What we want to know eventually is the scale of the deviation of
the cumulative number of individual word occurrences from the expected value,
the core question of this study.
The absolute size of the deviation depends on \(t_i\) as well as \(\tau\);
Eq.~(\ref{eq:expected_value_YS}) expresses that the cumulative number of word occurrences
increases more slowly with a larger \(t_i\), and therefore,
the size of the deviation of such words is supposed to be relatively smaller
than that of a smaller \(t_i\) if they use the same \(\tau\).
Thus, the size of the deviation should be normalized depending on \(t_i\)
using different values of \(\tau\).
Now we introduce a scale factor \(\lambda\) as follows:
\begin{align}
  s_i=t_i+\tau_i=\lambda t_i.
  \label{eq:omega}
\end{align}
Here \(\tau\), the observation period of the deviation, varies word by word,
and \(\lambda\) is constant for every word and greater than one by definition.
Substituting Eq.~(\ref{eq:omega}) into Eq.~(\ref{eq:expected_value_YS}), we obtain:
\begin{align}
  n^*_i(s_i)=\left(\frac{\lambda t_i}{t_i}\right)^{1-\alpha}=\lambda^{1-\alpha}\underset{\alpha\rightarrow 0}{=}\lambda.
  \label{eq:expected_value_omega}
\end{align}
This temporally normalized expected value of the cumulative number of word occurrences,
\(\lambda\), is used as a {\it reference} value
to measure the scale of the deviation for each word.
Replacing \(n\) in Eq.~(\ref{eq:pn_alpha0}) with \(x\lambda\),
that is, \(x\) times of the reference value, we obtain:
\begin{align}
  P\bigl(n_i(s_i&)=x\lambda\bigr)\nonumber\\
  &\underset{\alpha\rightarrow 0}{\sim}t_i(\lambda t_i)^{-x\lambda}\bigl\{(\lambda-1)t_i-x\lambda+2\bigr\}^{x\lambda-1}\nonumber\\
  &\hphantom{)}=\lambda^{-1}\left(1-\frac{1}{\lambda}-\frac{x-2/\lambda}{t_i}\right)^{x\lambda-1}.
  \label{eq:pn_alpha0_normalized}
\end{align}
The idea of \(x\), the scale of the deviation, is depicted in Fig.~{\ref{fig:diagram}}.
For a large majority of words, supposing \(t_i\gg x\sim 1\),
Eq.~(\ref{eq:pn_alpha0_normalized}) is approximated as:
\begin{align}
  P\bigl(n_i(s_i)=x\lambda\bigr)
  \underset{\alpha\rightarrow 0}{\sim}\frac{1}{\lambda-1}\left(1-\frac{1}{\lambda}\right)^{x\lambda},
  \label{eq:pn_alpha0_normalized_approx}
\end{align}
which is independent of \(t_i\), that is, independent of the word.
Hence, this formula represents the probability distribution of the fluctuation for all words.
This concise relationship is the other principal result of this article.
Equation (\ref{eq:pn_alpha0_normalized_approx}) clearly shows that the probability distribution
of the deviation scale decays exponentially.

We confirm that the general form (\ref{eq:pn_formal_approx}) and
the particular form for a sufficiently small \(\alpha\) (\ref{eq:pn_alpha0_normalized_approx})
well predict the actual behavior of the growth fluctuation in the cumulative number of word occurrences.

\begin{figure}
  \includegraphics[width=0.48\textwidth]{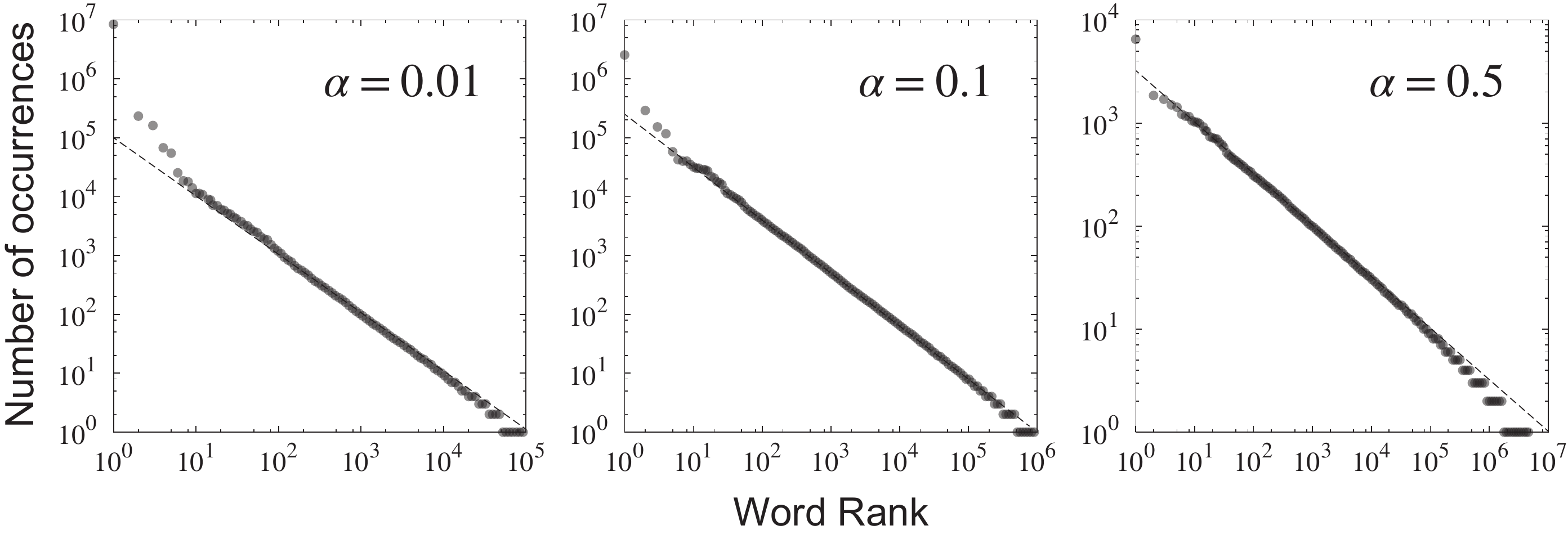}
  \caption{%
    The rank-frequency distribution.
    Black circles and dotted lines show the simulation results and theoretical curves
    proportional to \(\mbox{[word rank]}^{1-\alpha}\), respectively.
    }
  \label{fig:zipfs_law}
\end{figure}

\begin{figure}
  \includegraphics[width=0.48\textwidth]{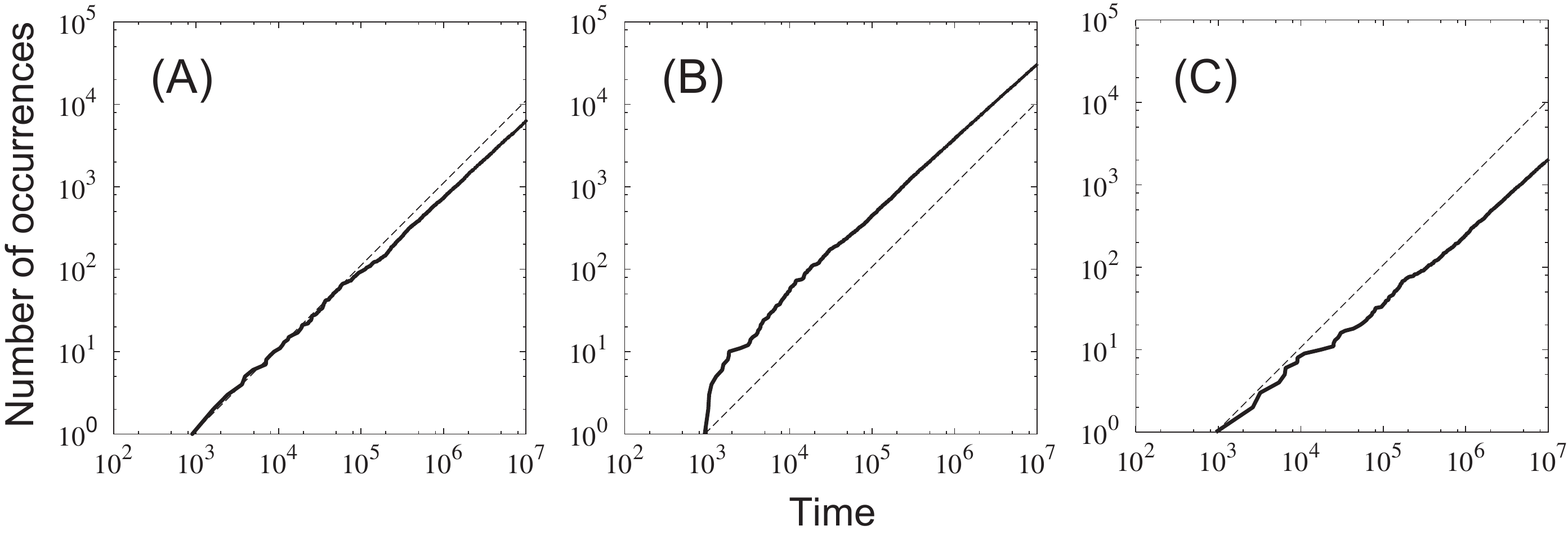}
  \caption{%
    The growth of the cumulative number of word occurrences of three sampled words
    created (A) 89th, (B) 90th, and (C) 91st in the case of \(\alpha=0.1\).
    Solid and dotted lines show the actual growth curves
    and the corresponding expected growth curves, respectively.
    }
  \label{fig:example}
\end{figure}

First, we ran the numerical simulation of the Yule--Simon process for different \(\alpha\) values of
0.01, 0.1, and 0.5, where the total number of word occurrences is \(10^7\);
consequently, the final vocabulary sizes are approximately \(10^5\), \(10^6\), and \(5\times 10^6\), respectively.
Figure \ref{fig:zipfs_law} shows the rank-frequency distribution for each \(\alpha\) value,
and we see that Zipf's law actually holds in every case with the power exponent \(1-\alpha\) predicted by the model.
We also show three typical patterns of the growth of the cumulative number of word occurrences,
especially in the case of \(\alpha=0.1\):
The word occurrence of the three sampled words (89th, 90th, and 91st)
increases (A) following, (B) exceeding, and (C) falling behind the expected growth curve,
respectively (Fig.~\ref{fig:example}).
These three words are created at a close time point,
however, exhibit differing growth courses.
This word-by-word fluctuation is what we have been trying to explain in this study.

\begin{figure*}
  \includegraphics[width=0.8\textwidth]{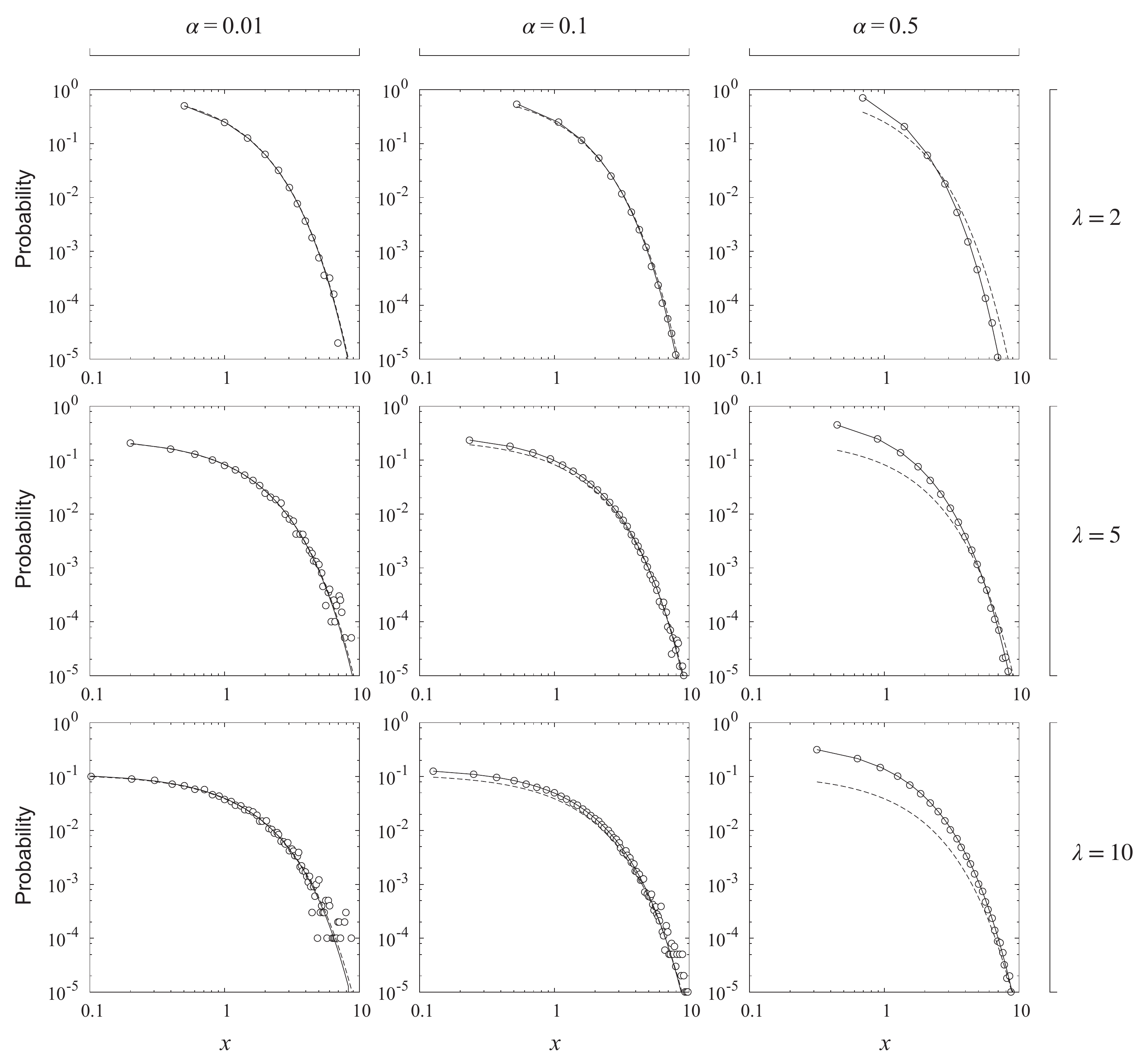}
  \caption{%
    Probability distributions of the deviation scale:
    the numerical results of the general form, Eq.~(\ref{eq:pn_formal_approx}) (solid lines);
    analytic curves drawn from the particular form for a sufficiently small \(\alpha\),
    Eq.~(\ref{eq:pn_alpha0_normalized_approx}) (dotted lines);
    and the numerical results of the simulation (white circles)
    for different \(\alpha\) and \(\lambda\) values.}
  \label{fig:comparison}
\end{figure*}

Using the simulation result, we measured the probability distribution of
the scale of the deviation from the reference value for different \(\lambda\) values of 2, 5, and 10.
To calculate the actual values of Eq.~(\ref{eq:pn_formal_approx}),
we used the same values of \(t_i\) in the simulation.
The results are shown in Fig.~\ref{fig:comparison};
for all \(\alpha\) and \(\lambda\) values, the simulation results exhibit a good match
with the general solution, and we conclude that our inductive derivation
of Eq.~(\ref{eq:pn_formal_approx}) and Eq.~(\ref{eq:pn_formal}) is valid.
In addition, we found that the simulation results exhibit good fit with
an exponential function with an identical characteristic scale of approximately 1;
the fitted parameters related to the simulation results are shown in Table \ref{tab:exp_params}.
This result seems to share the same relationship with Eq.~(\ref{eq:pn_alpha0_normalized_approx}),
which can be transformed into the form of \(\Delta\exp(-x)\) in the asymptotic limit of large \(\lambda\).
In this study, we keep the further discussion of this aspect on hold.
For small \(\alpha\) values, we also see a good match between the particular solution
and the other results; meanwhile, the mismatch between them increases for large \(\alpha\) values.
This is consistent with our assumption concerning asymptotic behavior of the particular solution.

\begin{table}[t]
  \caption{\label{tab:exp_params}%
    Fitted parameters for the simulation results for \((a\delta)\exp(-a(x-\delta))\),
    where \(\delta\) is a given interval in the calculation of the distribution function
    and corresponds to \(x\) values of the far-left white circles
    in Fig.~\ref{fig:comparison}.
  }
  \begin{ruledtabular}
    \begin{tabular}{lrcll}
      \multicolumn{1}{c}{\(\alpha\)}&
      \multicolumn{1}{c}{\(\lambda\)}&
      \multicolumn{1}{c}{\(\delta\)}&
      \multicolumn{1}{c}{\(a\)}&
      Std.~Err.\\
      \colrule
      0.01 & 2  & 0.501187 & 0.999232 & 0.0564\\
      0.01 & 5  & 0.199526 & 1.02095  & 0.01522\\
      0.01 & 10 & 0.102329 & 0.988124 & 0.008821\\
      0.1  & 2  & 0.524807 & 1.02668  & 0.05026\\
      0.1  & 5  & 0.234423 & 0.998086 & 0.01758\\
      0.1  & 10 & 0.125893 & 0.995484 & 0.008675\\
      0.5  & 2  & 0.691831 & 1.03491  & 0.07246\\
      0.5  & 5  & 0.446684 & 1.00143  & 0.04125\\
      0.5  & 10 & 0.316228 & 0.996059 & 0.02533\\
    \end{tabular}
  \end{ruledtabular}
\end{table}

In summary, we derived the probability distribution of the fluctuation
in the growth of the cumulative number of individual word occurrences
under the preferential attachment mechanism, based on the Yule--Simon process.
The distribution function was represented by the particular form for a sufficiently small \(\alpha\),
the creation rate of new vocabulary, that shows exponential decay with an increasing deviation scale.
We also obtained the general form of the probability distribution of word occurrences
and showed numerically that the solution follows the exponential decay in the growth fluctuation.
We confirmed that the theoretical solutions and the simulation results matched well,
concluding that our inductive derivation seems suitable.

The idea of the growth fluctuation in the preferential attachment dynamics
focused on this study and its solution raise further questions, as follows:
\begin{enumerate}
\item The BA graph was introduced to explain the growth of the web;
  do webpages or websites actually exhibit exponential decay in the fluctuation of their individual growth?

  The fact that only weak correlation between the size of a website and its age exists \cite{Adamic2002}
  implies the existence of significant individuality that might cause a deviation from the theoretical expectation.

\item Alternatively, do we find any phenomena that do not follow our result
  while showing the same population distribution, such as Zipf's law?

  This question is presumably related to the discussion on the scaling laws referred to
  previously \cite{Gabaix1999,Matia2005,Rybski2009}.
  In addition, it is a good reminder that
  incorporating the {\it fitness} function into the dynamics
  enables us to tune the individual growth rates;
  however, this distorts even the population distribution \cite{Bianconi2001}. 

\item Following from the previous question and based on Simon's derivation,
  which ensures the power-law population distribution,
  what form of the distribution function of the fluctuation
  can be derived if we use another rule in picking up a word from the class
  other than the uniformly random selection adopted here?
\end{enumerate}

We sincerely express our gratitude to
T.~Ikegami, M.~Oka, and K.~Sato for many fruitful discussions and suggestions.

\bibliography{manuscript}

\begin{thebibliography}{18}%
\makeatletter
\providecommand \@ifxundefined [1]{%
 \@ifx{#1\undefined}
}%
\providecommand \@ifnum [1]{%
 \ifnum #1\expandafter \@firstoftwo
 \else \expandafter \@secondoftwo
 \fi
}%
\providecommand \@ifx [1]{%
 \ifx #1\expandafter \@firstoftwo
 \else \expandafter \@secondoftwo
 \fi
}%
\providecommand \natexlab [1]{#1}%
\providecommand \enquote  [1]{``#1''}%
\providecommand \bibnamefont  [1]{#1}%
\providecommand \bibfnamefont [1]{#1}%
\providecommand \citenamefont [1]{#1}%
\providecommand \href@noop [0]{\@secondoftwo}%
\providecommand \href [0]{\begingroup \@sanitize@url \@href}%
\providecommand \@href[1]{\@@startlink{#1}\@@href}%
\providecommand \@@href[1]{\endgroup#1\@@endlink}%
\providecommand \@sanitize@url [0]{\catcode `\\12\catcode `\$12\catcode
  `\&12\catcode `\#12\catcode `\^12\catcode `\_12\catcode `\%12\relax}%
\providecommand \@@startlink[1]{}%
\providecommand \@@endlink[0]{}%
\providecommand \url  [0]{\begingroup\@sanitize@url \@url }%
\providecommand \@url [1]{\endgroup\@href {#1}{\urlprefix }}%
\providecommand \urlprefix  [0]{URL }%
\providecommand \Eprint [0]{\href }%
\providecommand \doibase [0]{http://dx.doi.org/}%
\providecommand \selectlanguage [0]{\@gobble}%
\providecommand \bibinfo  [0]{\@secondoftwo}%
\providecommand \bibfield  [0]{\@secondoftwo}%
\providecommand \translation [1]{[#1]}%
\providecommand \BibitemOpen [0]{}%
\providecommand \bibitemStop [0]{}%
\providecommand \bibitemNoStop [0]{.\EOS\space}%
\providecommand \EOS [0]{\spacefactor3000\relax}%
\providecommand \BibitemShut  [1]{\csname bibitem#1\endcsname}%
\let\auto@bib@innerbib\@empty
\bibitem [{\citenamefont {Yule}(1924)}]{Yule1924}%
  \BibitemOpen
  \bibfield  {author} {\bibinfo {author} {\bibfnamefont {G.~U.}\ \bibnamefont
  {Yule}},\ }\href@noop {} {\bibfield  {journal} {\bibinfo  {journal}
  {Phil.~Trans.~Roy.~Soc.~B}\ }\textbf {\bibinfo {volume} {213}},\ \bibinfo
  {pages} {21} (\bibinfo {year} {1924})}\BibitemShut {NoStop}%
\bibitem [{\citenamefont {Baca\"{e}r}(2011)}]{Bacaer2011}%
  \BibitemOpen
  \bibfield  {author} {\bibinfo {author} {\bibfnamefont {N.}~\bibnamefont
  {Baca\"{e}r}},\ }\enquote {\bibinfo {title} {Yule and evolution (1924)},}\
  in\ \href@noop {} {\emph {\bibinfo {booktitle} {A Short History of
  Mathematical Population Dynamics}}}\ (\bibinfo  {publisher} {Springer-Verlag,
  London},\ \bibinfo {year} {2011})\ pp.\ \bibinfo {pages} {81--88}\BibitemShut
  {NoStop}%
\bibitem [{\citenamefont {Simkin}\ and\ \citenamefont
  {Roychowdhury}(2011)}]{Simkin2011}%
  \BibitemOpen
  \bibfield  {author} {\bibinfo {author} {\bibfnamefont {M.~V.}\ \bibnamefont
  {Simkin}}\ and\ \bibinfo {author} {\bibfnamefont {V.~P.}\ \bibnamefont
  {Roychowdhury}},\ }\href@noop {} {\bibfield  {journal} {\bibinfo  {journal}
  {Physics Reports}\ }\textbf {\bibinfo {volume} {502}},\ \bibinfo {pages} {1}
  (\bibinfo {year} {2011})}\BibitemShut {NoStop}%
\bibitem [{\citenamefont {Simon}(1955)}]{Simon1955}%
  \BibitemOpen
  \bibfield  {author} {\bibinfo {author} {\bibfnamefont {H.~A.}\ \bibnamefont
  {Simon}},\ }\href@noop {} {\bibfield  {journal} {\bibinfo  {journal}
  {Biometrika}\ }\textbf {\bibinfo {volume} {42}},\ \bibinfo {pages} {425}
  (\bibinfo {year} {1955})}\BibitemShut {NoStop}%
\bibitem [{\citenamefont {Zipf}(1935)}]{Zipf1935}%
  \BibitemOpen
  \bibfield  {author} {\bibinfo {author} {\bibfnamefont {G.~K.}\ \bibnamefont
  {Zipf}},\ }\href@noop {} {\emph {\bibinfo {title} {The Psycho-Biology of
  Language}}}\ (\bibinfo  {publisher} {Houghton Mifflin Company, Boston},\
  \bibinfo {year} {1935})\BibitemShut {NoStop}%
\bibitem [{\citenamefont {Mahmoud}(2008)}]{Mahmoud2008}%
  \BibitemOpen
  \bibfield  {author} {\bibinfo {author} {\bibfnamefont {H.}~\bibnamefont
  {Mahmoud}},\ }\href@noop {} {\emph {\bibinfo {title} {P\'{o}lya Urn
  Models}}}\ (\bibinfo  {publisher} {Chapman and Hall/CRC},\ \bibinfo {year}
  {2008})\BibitemShut {NoStop}%
\bibitem [{\citenamefont {Barab\'{a}si}\ \emph {et~al.}(1999)\citenamefont
  {Barab\'{a}si}, \citenamefont {Albert},\ and\ \citenamefont
  {Jeong}}]{Barabasi1999}%
  \BibitemOpen
  \bibfield  {author} {\bibinfo {author} {\bibfnamefont {A.-l.}\ \bibnamefont
  {Barab\'{a}si}}, \bibinfo {author} {\bibfnamefont {R.}~\bibnamefont
  {Albert}}, \ and\ \bibinfo {author} {\bibfnamefont {H.}~\bibnamefont
  {Jeong}},\ }\href@noop {} {\bibfield  {journal} {\bibinfo  {journal} {Physica
  A}\ }\textbf {\bibinfo {volume} {272}},\ \bibinfo {pages} {173^^e2^^80^^93}
  (\bibinfo {year} {1999})}\BibitemShut {NoStop}%
\bibitem [{\citenamefont {Bornholdt}\ and\ \citenamefont
  {Ebel}(2001)}]{Bornholdt2001}%
  \BibitemOpen
  \bibfield  {author} {\bibinfo {author} {\bibfnamefont {S.}~\bibnamefont
  {Bornholdt}}\ and\ \bibinfo {author} {\bibfnamefont {H.}~\bibnamefont
  {Ebel}},\ }\href@noop {} {\bibfield  {journal} {\bibinfo  {journal}
  {Phys.~Rev.~E}\ }\textbf {\bibinfo {volume} {64}},\ \bibinfo {pages} {035104}
  (\bibinfo {year} {2001})}\BibitemShut {NoStop}%
\bibitem [{\citenamefont {Bianconi}\ and\ \citenamefont
  {Barab\'{a}si}(2001)}]{Bianconi2001}%
  \BibitemOpen
  \bibfield  {author} {\bibinfo {author} {\bibfnamefont {G.}~\bibnamefont
  {Bianconi}}\ and\ \bibinfo {author} {\bibfnamefont {A.-l.}\ \bibnamefont
  {Barab\'{a}si}},\ }\href@noop {} {\bibfield  {journal} {\bibinfo  {journal}
  {Europhys.~Lett.}\ }\textbf {\bibinfo {volume} {54}},\ \bibinfo {pages} {436}
  (\bibinfo {year} {2001})}\BibitemShut {NoStop}%
\bibitem [{\citenamefont {Krapivsky}\ and\ \citenamefont
  {Redner}(2002{\natexlab{a}})}]{Krapivsky2002b}%
  \BibitemOpen
  \bibfield  {author} {\bibinfo {author} {\bibfnamefont {P.~L.}\ \bibnamefont
  {Krapivsky}}\ and\ \bibinfo {author} {\bibfnamefont {S.}~\bibnamefont
  {Redner}},\ }\href@noop {} {\bibfield  {journal} {\bibinfo  {journal}
  {Phys.~Rev.~Lett.}\ }\textbf {\bibinfo {volume} {89}},\ \bibinfo {pages}
  {258703} (\bibinfo {year} {2002}{\natexlab{a}})}\BibitemShut {NoStop}%
\bibitem [{\citenamefont {Cattuto}\ \emph {et~al.}(2009)\citenamefont
  {Cattuto}, \citenamefont {Barrat}, \citenamefont {Baldassarri}, \citenamefont
  {Schehr},\ and\ \citenamefont {Loreto}}]{Cattuto2009}%
  \BibitemOpen
  \bibfield  {author} {\bibinfo {author} {\bibfnamefont {C.}~\bibnamefont
  {Cattuto}}, \bibinfo {author} {\bibfnamefont {A.}~\bibnamefont {Barrat}},
  \bibinfo {author} {\bibfnamefont {A.}~\bibnamefont {Baldassarri}}, \bibinfo
  {author} {\bibfnamefont {G.}~\bibnamefont {Schehr}}, \ and\ \bibinfo {author}
  {\bibfnamefont {V.}~\bibnamefont {Loreto}},\ }\href@noop {} {\bibfield
  {journal} {\bibinfo  {journal} {PNAS}\ }\textbf {\bibinfo {volume} {106}},\
  \bibinfo {pages} {10511} (\bibinfo {year} {2009})}\BibitemShut {NoStop}%
\bibitem [{\citenamefont {Gupta}\ \emph {et~al.}(2010)\citenamefont {Gupta},
  \citenamefont {Li}, \citenamefont {Yin},\ and\ \citenamefont
  {Han}}]{Gupta2010}%
  \BibitemOpen
  \bibfield  {author} {\bibinfo {author} {\bibfnamefont {M.}~\bibnamefont
  {Gupta}}, \bibinfo {author} {\bibfnamefont {R.}~\bibnamefont {Li}}, \bibinfo
  {author} {\bibfnamefont {Z.}~\bibnamefont {Yin}}, \ and\ \bibinfo {author}
  {\bibfnamefont {J.}~\bibnamefont {Han}},\ }\href {\doibase
  10.1145/1882471.1882480} {\bibfield  {journal} {\bibinfo  {journal} {SIGKDD
  Explor. Newsl.}\ }\textbf {\bibinfo {volume} {12}},\ \bibinfo {pages} {58}
  (\bibinfo {year} {2010})}\BibitemShut {NoStop}%
\bibitem [{\citenamefont {Albert}\ and\ \citenamefont
  {Barab\'{a}si}(2002)}]{Albert2002}%
  \BibitemOpen
  \bibfield  {author} {\bibinfo {author} {\bibfnamefont {R.}~\bibnamefont
  {Albert}}\ and\ \bibinfo {author} {\bibfnamefont {A.-l.}\ \bibnamefont
  {Barab\'{a}si}},\ }\href@noop {} {\bibfield  {journal} {\bibinfo  {journal}
  {Rev.~Mod.~Phys.}\ }\textbf {\bibinfo {volume} {74}},\ \bibinfo {pages}
  {47^^e2^^80^^93} (\bibinfo {year} {2002})}\BibitemShut {NoStop}%
\bibitem [{\citenamefont {Krapivsky}\ and\ \citenamefont
  {Redner}(2002{\natexlab{b}})}]{Krapivsky2002a}%
  \BibitemOpen
  \bibfield  {author} {\bibinfo {author} {\bibfnamefont {P.~L.}\ \bibnamefont
  {Krapivsky}}\ and\ \bibinfo {author} {\bibfnamefont {S.}~\bibnamefont
  {Redner}},\ }\href@noop {} {\bibfield  {journal} {\bibinfo  {journal}
  {J.~Phys.~A: Math.~Gen.}\ }\textbf {\bibinfo {volume} {35}},\ \bibinfo
  {pages} {9517} (\bibinfo {year} {2002}{\natexlab{b}})}\BibitemShut {NoStop}%
\bibitem [{\citenamefont {Gabaix}(1999)}]{Gabaix1999}%
  \BibitemOpen
  \bibfield  {author} {\bibinfo {author} {\bibfnamefont {X.}~\bibnamefont
  {Gabaix}},\ }\href@noop {} {\bibfield  {journal} {\bibinfo  {journal}
  {Quart.~J.~Econ.}\ }\textbf {\bibinfo {volume} {114}},\ \bibinfo {pages}
  {739} (\bibinfo {year} {1999})}\BibitemShut {NoStop}%
\bibitem [{\citenamefont {Matia}\ \emph {et~al.}(2005)\citenamefont {Matia},
  \citenamefont {Nunes~Amaral}, \citenamefont {Luwel}, \citenamefont {Moed},\
  and\ \citenamefont {Stanley}}]{Matia2005}%
  \BibitemOpen
  \bibfield  {author} {\bibinfo {author} {\bibfnamefont {K.}~\bibnamefont
  {Matia}}, \bibinfo {author} {\bibfnamefont {L.~A.}\ \bibnamefont
  {Nunes~Amaral}}, \bibinfo {author} {\bibfnamefont {M.}~\bibnamefont {Luwel}},
  \bibinfo {author} {\bibfnamefont {H.~F.}\ \bibnamefont {Moed}}, \ and\
  \bibinfo {author} {\bibfnamefont {H.~E.}\ \bibnamefont {Stanley}},\ }\href
  {\doibase 10.1002/asi.v56:9} {\bibfield  {journal} {\bibinfo  {journal}
  {J.~Am.~Soc.~Inf.~Sci.~Technol.}\ }\textbf {\bibinfo {volume} {56}},\
  \bibinfo {pages} {893} (\bibinfo {year} {2005})}\BibitemShut {NoStop}%
\bibitem [{\citenamefont {Rybski}\ \emph {et~al.}(2009)\citenamefont {Rybski},
  \citenamefont {Buldyrev}, \citenamefont {Havlin}, \citenamefont {Liljeros},\
  and\ \citenamefont {Makse}}]{Rybski2009}%
  \BibitemOpen
  \bibfield  {author} {\bibinfo {author} {\bibfnamefont {D.}~\bibnamefont
  {Rybski}}, \bibinfo {author} {\bibfnamefont {S.~V.}\ \bibnamefont
  {Buldyrev}}, \bibinfo {author} {\bibfnamefont {S.}~\bibnamefont {Havlin}},
  \bibinfo {author} {\bibfnamefont {F.}~\bibnamefont {Liljeros}}, \ and\
  \bibinfo {author} {\bibfnamefont {H.~A.}\ \bibnamefont {Makse}},\ }\href@noop
  {} {\bibfield  {journal} {\bibinfo  {journal} {PNAS}\ }\textbf {\bibinfo
  {volume} {106}},\ \bibinfo {pages} {12640} (\bibinfo {year}
  {2009})}\BibitemShut {NoStop}%
\bibitem [{\citenamefont {Adamic}\ and\ \citenamefont
  {Huberman}(2002)}]{Adamic2002}%
  \BibitemOpen
  \bibfield  {author} {\bibinfo {author} {\bibfnamefont {L.~A.}\ \bibnamefont
  {Adamic}}\ and\ \bibinfo {author} {\bibfnamefont {B.~A.}\ \bibnamefont
  {Huberman}},\ }\href@noop {} {\bibfield  {journal} {\bibinfo  {journal}
  {Glottometrics}\ }\textbf {\bibinfo {volume} {3}},\ \bibinfo {pages} {143}
  (\bibinfo {year} {2002})}\BibitemShut {NoStop}%
\end{thebibliography}%

\end{document}